\documentclass[fleqn,usenatbib]{mnras}

\usepackage{newtxtext,newtxmath}

\usepackage[T1]{fontenc}
\usepackage{ae,aecompl}

\usepackage{graphicx}	
\usepackage{amsmath}	
\usepackage{amssymb}	
\usepackage{subfigure}
\usepackage{soul}

\title[]{Differential Astrometric Framework for the Jupiter Relativistic
Experiment with Gaia}

\author[U. Abbas et al.]{
Ummi Abbas,$^{1}$\thanks{E-mail: ummi.abbas@inaf.it (UA)}
Beatrice Bucciarelli,$^{1}$
Mario G. Lattanzi$^{1}$
\\
$^{1}$INAF - Osservatorio Astrofisico di Torino, Via Osservatorio 20, Pino Torinese, Italy 10025
}

\date{Accepted 2019 January 30. Received 2019 January 25; in original form 2018 December 28.}

\pubyear{2018}

\begin{document}
\label{firstpage}
\pagerange{\pageref{firstpage}--\pageref{lastpage}}
\maketitle

\begin{abstract}
We employ differential astrometric methods to establish a small field reference frame stable at the micro-arcsecond ($\mu$as) level on short timescales 
using high-cadence simulated observations taken by Gaia in February 2017 of a bright star close to the limb of Jupiter, as part of the relativistic experiment on Jupiter's quadrupole.
We achieve sub$\mu$as-level precision along scan through a suitable transformation of the 
field angles into a small-field tangent plane and a least-squares fit over 
several overlapping frames for estimating the plate and geometric calibration parameters with tens of reference stars that lie 
within $\sim$0.5 degs from the target star, assuming perfect knowledge of stellar proper motions and parallaxes. 
Furthermore, we study the effects of unmodeled astrometric parameters on the residuals
and find that proper motions have a stronger effect than 
unmodeled parallaxes. For e.g., unmodeled Gaia DR2 proper motions introduce extra
residuals of $\sim$23$\mu$as (AL) and 69$\mu$as (AC) versus the $\sim$5$\mu$as (AL) and 17$\mu$as (AC) due to unmodeled parallaxes.
On the other hand, assuming catalog errors in 
the proper motions such as those from Gaia DR2 has a minimal impact on the 
stability introducing sub$\mu$as and $\mu$as level residuals in the along and across scanning direction, respectively. Finally, the effect of a coarse knowledge in the satellite velocity components (with time dependent errors of 10$\mu$as) is capable of enlarging the size of the residuals to roughly 0.2 mas.

\end{abstract}




\section{Introduction}

The on-going Gaia space mission \citep{prusti2016} has provided high-precision global astrometric catalogues in two successive data releases over the past few years \citep{brown2016, brown2018}.
This has heralded the beginning of an era of high-precision astrometry in which the models and methodologies are required to demonstrate at least micro-arcsecond ($\mu$as) level, if not higher, precision \citep{lattanzi2012}.
Indeed, as part of the Gaia Data Processing and Analysis Consortium (DPAC) activities, the description of stellar positions in Gaia measurements, using the baseline Gaia RElativistic Model (GREM) model in the Astrometric Global Iterative Solution (AGIS) \citep{klioner2003} or its Relativistic Astrometric MODel (RAMOD) \citep{crosta2008, crosta2010} counterpart in the Global Sphere Reconstruction (GSR) module for the astrometric verification of the AGIS solution, is accurate at the sub-$\mu$as level. 

The astrometric parameters (position, proper motion and parallaxes) as provided in past Gaia data releases are absolute parameters obtained by the AGIS solution, (e.g. \citealt{lindegren2016, lindegren2018}). The scanning motion 
of the satellite is specifically designed to provide measurements over the whole celestial sphere 
in 6 months' time through a combination of three motion: a) rotation of Gaia around its spin axis every 6 hours, b) the precession of the spin axis every 63 days, and, c) the orbital motion of the satellite around the Sun \citep{lindegren2016}. It is possible to perform differential astrometry starting with Gaia's global astrometric measurements, constructing a high-precision
inertial astrometric reference frame over small fields with a size essentially defined by the dimensions of 
Gaia's field of view (FOV). This approach requires a sophisticated methodology that uses the set of coordinates (field angles) in the focal plane of the satellite as input and depends mainly on the geometric calibrations of the CCDs and the satellite attitude. 
The main challenge to face concerns the ability to robustly define such a local reference frames over the very different time-scales associated to diverse scientific applications of this approach, such as relativistic deflection of light tests (duration of a few days), astrometric microlensing events (duration up to monhts), and orbital reconstruction of binary and substellar companions, including exoplanets (duration of years). The developed methodology must be able to account for highly different distributions and number of comparison stars at different epochs of Gaia observations (transits), depending on the satellite's scanning direction, as well as highly changing geometric calibrations, particularly when the timespan of the observations exceeds a few weeks \citep{lindegren2018}. 

In \citealt{abbas2017} we provided an initial assessment of this approach using Gaia simulated high-cadence observations around the ecliptic poles as a backbone, finding that the so-defined local reference frame would be stable at the $\mu$as level with a modest number (30-40) of stars over a $0.24\times0.24$ deg field. In this paper we further develop our methodology, focusing on the observational scenario setup for the purpose of the Gaia relativistic light deflection experiment around Jupiter, particularly due to its flattened mass distribution that induces a quadrupole term (hence the name GAREQ - Gaia Relativistic Experiment on Jupiter's Quadrupole, \citealt{crosta2006, crosta2008a, crosta2008b}). In order to carry out the experiment, the Gaia nominal scanning law was specifically optimized in its two free parameters, the initial precession and the spin phase angles, in order to obtain the most
suitable arrangement of a bright star observed as close as possible to Jupiter's limb \citep{prusti2016, deBruijne2010}. 
As the quadrupole light deflection term scales with the inverse cube of the impact parameter, the induced effect is transient and lasts only about 20 hours for a detection at the 10$\mu$as level (for a light ray grazing Jupiter's limb the effect is $\sim$240 $\mu$as). An optimized observational campaign was therefore key to obtaining the best possible signal as a compromise with the dominant stray light effects very close to Jupiter and away from the bright trail of light left behind by Jupiter on the CCDs during the scanning motion of Gaia. Furthermore, the GAREQ experiment is a special case scenario, with high-cadence observations obtained by Gaia on several successive transits on a timespan of 3 days. 

In this paper we perform a differential astrometric analysis of Gaia simulated observations of a GAREQ event that occurred in February 2017, to gauge the best approach towards obtaining a reference frame stable at the $\mu$as level.
The analysis described here improves on \citealt{abbas2017} in several respects, including 
a) a novel implementation of the methodology to significantly reduce the number of unknown parameters, b) the inclusion of stellar parallaxes, and c) the investigation of the effects on the stability of the  reference frame in the presence of unmodeled proper motions and parallaxes as well as when the known astrometric parameters are removed a priori. 

The paper is laid out as follows: in Sec. \ref{sect:simul} we discuss the simulation setup, in Sec. \ref{sect:transforms} we present the procedure for transforming the field angles into the tangent plane and then for obtaining a least squares fit on the unknown parameters. In the following Sec. \ref{results} we present our most relevant findings, closing up with a summary and brief discussion in Sec. \ref{discussion}.

\section{The Gaia simulations for GAREQ}\label{sect:simul}
We will attempt to construct a small field ($\sim$0.3 sq. degs) reference frame using the real-time 
observations with Gaia of the field angles as measured in its FOV. These field angles are given by $\eta$ and $\zeta$ along and across Gaia's scanning direction respectively. 
The angles are simulated with the AGISLab software package that is ideal for running realistic 
small-scale experiments on a laptop and is faithful to the Gaia satellite. The package uses a subset 
of the most important functionalities of the AGIS mainstream pipeline used to analyze the actual Gaia data \citep{holl2012a}.

\subsection{The scanning law and its optimization}\label{sect:scanlaw}

\begin{figure}
    \centering
    \includegraphics{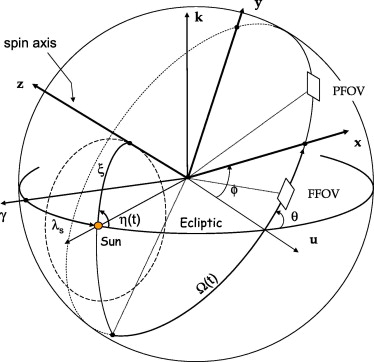}
    \caption{The scanning law of Gaia demonstrating the precession of the spin axis around the Sun,
    the solar aspect angle of $45^\circ$, and the two heliotropic angles: the precession 
    and spin phases given by $\eta(t)$ and $\Omega(t)$ respectively. (Image courtesy: \citealt{mignard2011})}
    \label{fig:scan_law}
\end{figure}

The Gaia space satellite scans the sky through a uniform revolving motion that is designed to 
maximize the uniformity of the coverage with the following three main ingredients \citep{prusti2016}: 
a) spin rate, $\omega_z \sim$ 60.0"$s^{-1}$ around the spin axis corresponding to a rotation
period of 6 hours, b) solar aspect angle, $\xi$ = 45$^\circ$ to maintain a good thermal stability (and in turn a constant basic angle between the two fields of view) and high parallax sensitivity that depends on $\sin{\xi}$, and,
c) precession of the spin axis around the Sun with a period of 63 days that causes a sinusoidal 
varying across-scan speed over 6 hours of the stellar images across the focal plane. A figure showing these three motions is given in Fig.~\ref{fig:scan_law}.

\begin{figure*}
    \hfill
    \subfigure[]{\includegraphics[width=8.5cm]{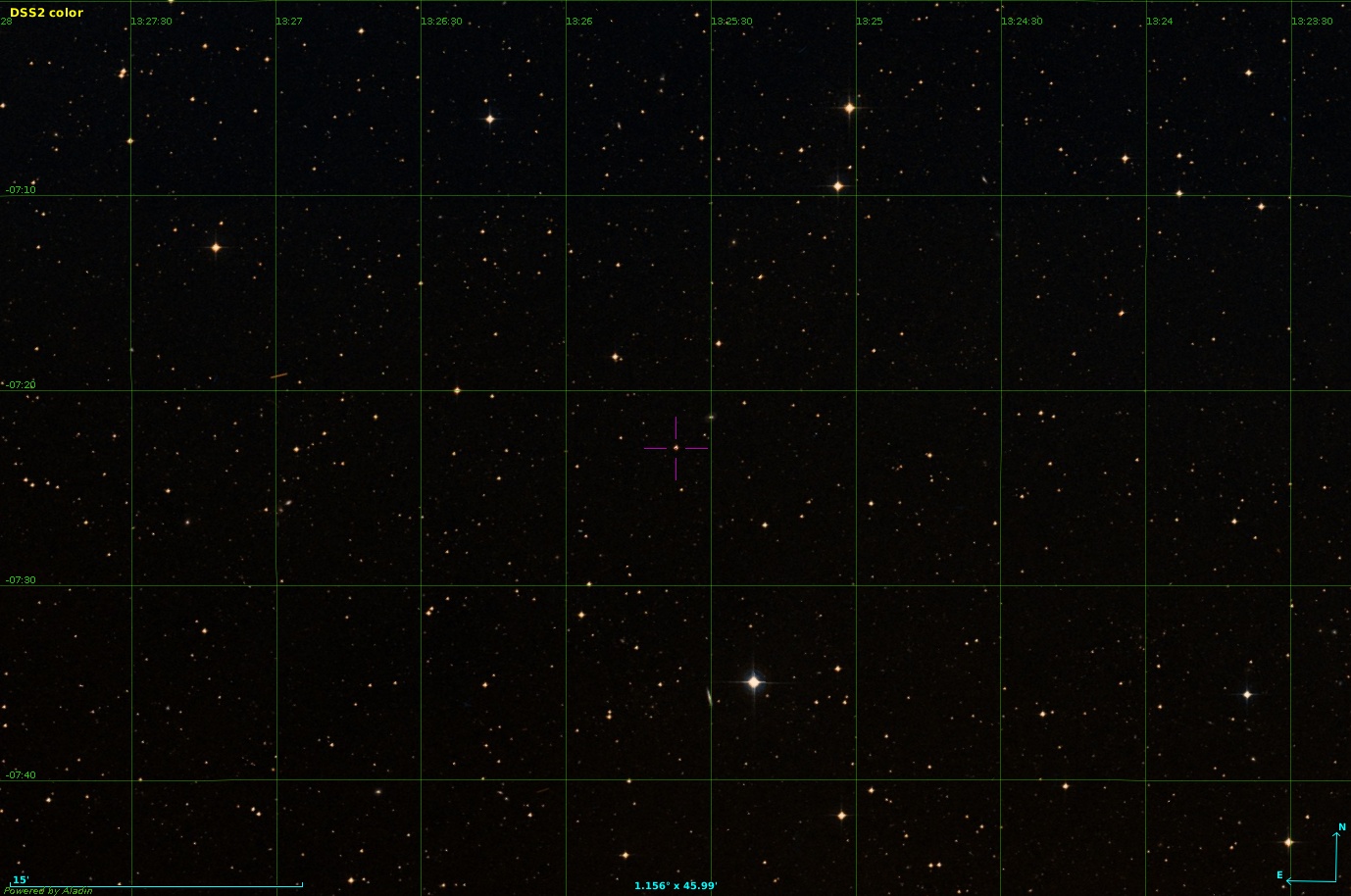}}
    \subfigure[]{\includegraphics[width=8.5cm]{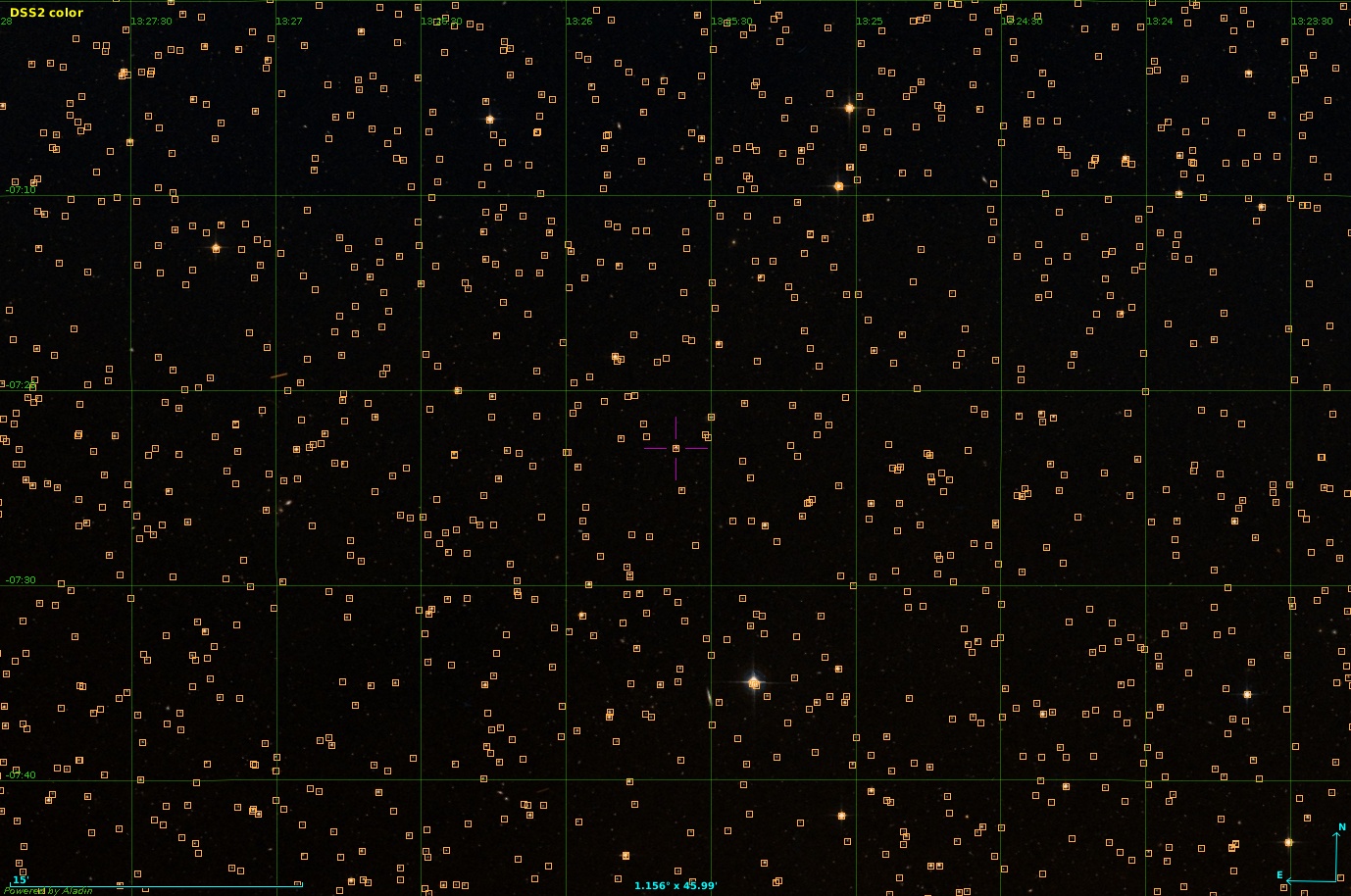}}
    \caption{The star field surrounding the target star located at $\alpha$ = 201.4053, $\delta$ = 
    -7.3830. Left panel (a) shows the DSS image of the field of size 1.156 degs x 45.99' and the right panel (b) shows the Gaia DR2 stars overlayed. (courtesy: Aladin)}
    \label{fig:star_field}
\end{figure*}

This uniform motion of the satellite unfolds in the ecliptic plane with the Sun at its center and
is determined by two heliotropic angles: a) the precession phase, $\nu(t)$ that is the angle between the ecliptic plane and the plane (Sun-z) containing the Sun and satellite $z$ axis, 
b) the spin phase, $\Omega(t)$ that is the angle on the great circle between the Sun-z and the satellite $zx$ planes
(see Fig.\ref{fig:scan_law}, note that in the figure $\nu(t)$ is given as $\eta(t)$).
Their time dependent equations have two free parameters; the initial precession phase and the initial spin phase angles at the start of science operations \citep{prusti2016}. These free parameters 
have been optimized in order to obtain the best possible observing conditions necessary to obtain the
highest possible signal for the GAREQ experiment. A dedicated and non-trivial optimization campaign carried out by 
the RElativistic Modeling And Testing (REMAT) group within the Gaia Data Processing
and Analysis Consortium was successful in predicting a set of 3 bright stars ($G = 15.75$ mag) all observed close to Jupiter with the brightest star ($G = 12.7$ mag) seen barely a few
arcseconds from Jupiter's limb during its passage across the sky 
\citep{klioner2014a, klioner2014b}.
This star field is shown in Fig.\ref{fig:star_field} where each observing time is clocked when the 
stellar image centroid passes the fiducial line of a CCD and represents the fundamental observed quantity. 
The fiducial line is generally halfway between the first and last line used during the time-delayed
integration (TDI) mode of Gaia. In the Focal Plane of Gaia there are 9 fiducial lines for all the
columns that make up the Astrometric Field having 62 CCDs from 7 rows and 9 columns, except for
the last column with 6 CCD rows (Prusti et al. 2016, Fig. 4).

\subsection{The simulated field angles and reference stars}\label{sect:refstars}

The adopted setup is based on the Gaia satellite equipped with two FOVs that are separated by the 
basic angle of 106.5 degs and rotating at a fixed rate of 59".9605 $s^{-1}$ around its spin axis
(Gaia Collaboration et al. 2016). The simulator is run using nominal CCD size, focal plane geometry and FOV size and orbital parameters. The observed source direction or its proper direction, is computed using a suitable relativistic model necessary for highly accurate astrometric observations in the parametrized post-Newtonian (PPN) framework and specifically adopted by Gaia \citep{klioner2003}. 
This formulation takes into account the relativistic modeling of the motion of the observer and modeling of relativistic aberration and gravitational light deflection, as well as a relativistic treatment of the parallaxes and proper motions. 

\begin{figure*}
    \subfigure[]{\includegraphics[width=8.5cm]{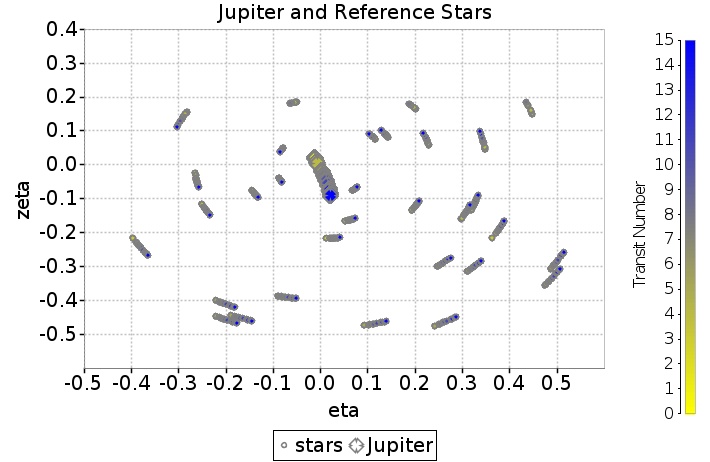}}
    \subfigure[]{\includegraphics[width=8.5cm]{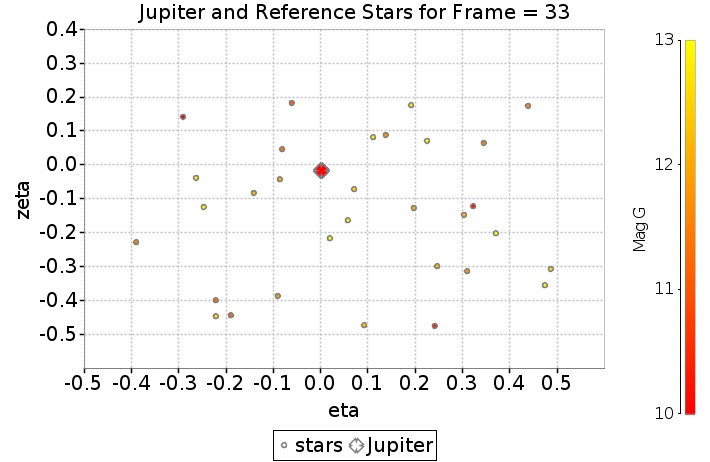}}
    \subfigure[]{\includegraphics[width=8.5cm]{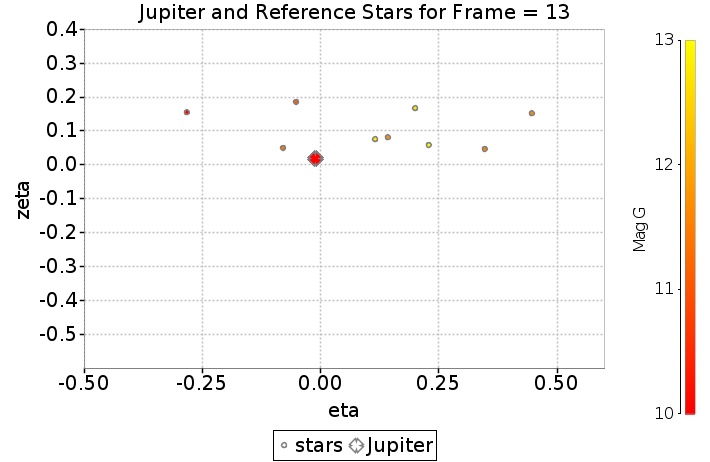}}
    \subfigure[]{\includegraphics[width=8.5cm]{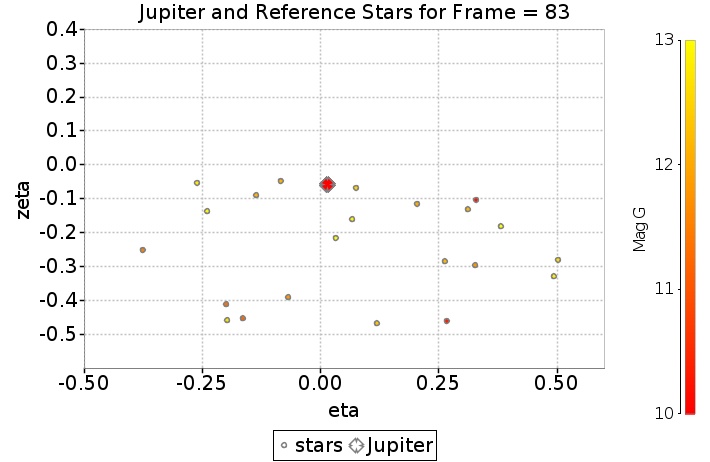}}
    \caption{The simulated star field surrounding the target star (not shown here) and Jupiter's position on various
    transits as seen by Gaia. In reality, Jupiter will not be observed in the AF due to its extreme brightness 
    (V=-2.7 mag).
    The top left panel shows the reference stars and Jupiter as seen on all
    transits, colour coded by the transit number.
    The top right panel (b) shows the reference frame with the maximum number of reference stars seen over the 
    15 transits. The symbols are colour coded according to their magnitude as shown in the color bar at the right. 
    The small circles show the reference stars whereas the larger diamond denotes Jupiter. 
    Lower left panel (c) shows the reference stars as seen in frame 13, and (d) shows the stars as seen in frame 83.}
    \label{fig:frames}
\end{figure*}

The first 15 transits around the target star (G=12.68 mag) in Februrary 2017 were used
that also coincide with when Jupiter is seen within the same frame. 
In order to maximize the number of observations, close bright stars (10 $<$ G $<$ 13 mag) having 2d windows, and, 
hence the $\eta$ and $\zeta$ coordinates, are chosen to lie within $\pm$30 secs of the observing time of the target 
star for the same CCD column.
This ensures that the target star always lies at the center of the frame surrounded by
the set of reference stars that are not necessarily seen by Gaia on each and every one of the 15 
transits during which the target star is present, mainly depending on the scanning direction during those instants of time causing different orientations of the astrometric field. This leads to different 'subsets' of reference
stars observed on different transits as some areas of the field are 'cut off' depending on the scanning angle, for 
e.g.
the target star that is ideally seen in the central row 4 leading to the whole surrounding star field being visible,
could be seen on row 1 or 7 in the following transit caused by the shift in the focal plane leading to the whole 
lower 
or upper part respectively of the star field to remain hidden from Gaia and unobserved (see Fig.\ref{fig:frames}).
Under these conditions we have chosen the reference frame to contain the maximum number of reference stars, 
in this case 31 stars, surrounding the target star. This high number of stars are given by the observing times as 
seen by AF1 during Gaia's fifth transit over the star field and converted into spatial coordinates as described 
in Section\ref{sect:alacrate}. 
The AF1-AF8 observations were used, AF9 observations were rejected due to the presence of the WFS. 
Table \ref{table:1} shows the total number of times each of the reference stars is seen. All the reference stars are seen 8 times during a whole transit except for star no. 21 that is seen a few times less on the second transit as it falls at the very edge of the FOV as mentioned earlier.
Table \ref{table:2} gives the total number of reference stars seen on each transit that 
constitutes 8 frames.
\begin{table*}
\centering
\begin{tabular}{| c c c c c c c c |} 
 \hline 
 starId & nObs & $\pi$[mas] & $\sigma_\pi$[mas] & $\mu_{\alpha*}$[mas/yr] & $\sigma_{\mu_{\alpha*}}$[mas/yr] & $\mu_\delta$[mas/yr] & $\sigma_{\mu_\delta}$[mas/yr] \\ 
 \hline
1 & 104 & 0.580 & 0.047 & -43.699 & 0.079 & -15.470 & 0.057\\
2 & 88 & 0.271 & 0.037 & -5.453 & 0.067 & -0.527 & 0.052\\
3 & 80 & 2.325 & 0.051 & -26.436 & 0.079 & 6.346 & 0.070\\
4 & 80 & 3.059 & 0.051 & -43.430 & 0.090 & -3.309 & 0.067\\
5 & 88 & 1.008 & 0.050 & -2.197 & 0.066 & 2.964 & 0.053\\
6 & 104 & 0.689 & 0.066 & 0.590 & 0.081 & -9.552 & 0.077\\
7 & 88 & 2.810 & 0.058 & 20.508 & 0.098 & -49.644 & 0.117\\
8 & 96 & 1.866 & 0.039 & -19.363 & 0.077 & -10.430 & 0.056\\
9 & 56 & 0.876 & 0.065 & 4.522 & 0.236 & -8.638 & 0.096\\
10 & 112 & 0.417 & 0.037 & -5.592 & 0.080 & 0.653 & 0.055\\
11 & 112 & 6.977 & 0.040 & -76.545 & 0.074 & -7.742 & 0.060\\
12 & 112 & 1.179 & 0.043 & -8.800 & 0.077 & 12.454 & 0.059\\
13 & 104 & 2.350 & 0.036 & -16.248 & 0.076 & 15.734 & 0.048\\
14 & 80 & 3.980 & 0.045 & 22.883 & 0.083 & -17.226 & 0.058\\
15 & 80 & 3.443 & 0.050 & -22.019 & 0.080 & -18.794 & 0.069\\
16 & 88 & 5.450 & 0.045 & -23.764 & 0.081 & -16.905 & 0.063\\
17 & 88 & 5.609 & 0.049 & -112.627 & 0.081 & -30.181 & 0.083\\
18 & 88 & 2.569 & 0.038 & -29.627 & 0.082 & 0.251 & 0.057\\
19 & 88 & 8.986 & 0.078 & 104.889 & 0.104 & -48.145 & 0.102\\
20 & 88 & 1.297 & 0.046 & -10.022 & 0.075 & 4.051 & 0.055\\
21 & 85 & 0.556 & 0.045 & -25.758 & 0.086 & -7.016 & 0.064\\
22 & 112 & 1.314 & 0.040 & -5.323 & 0.070 & -0.234 & 0.053\\
23 & 96 & 0.603 & 0.046 & -10.050 & 0.086 & -24.174 & 0.077\\
24 & 104 & 1.401 & 0.038 & -16.193 & 0.072 & -4.848 & 0.056\\
25 & 80 & 1.534 & 0.049 & -0.476 & 0.075 & -3.760 & 0.059\\
26 & 80 & 0.671 & 0.094 & -3.727 & 0.209 & 2.446 & 0.272\\
27 & 80 & 2.643 & 0.101 & 4.523 & 0.166 & -7.290 & 0.130\\
28 & 104 & 1.316 & 0.066 & 4.191 & 0.094 & 0.802 & 0.071\\
29 & 56 & 0.351 & 0.040 & 5.266 & 0.079 & -15.818 & 0.065\\
30 & 88 & 2.628 & 0.072 & -31.088 & 0.117 & -5.173 & 0.109\\
31 & 56 & 3.439 & 0.053 & -57.511 & 0.116 & -22.168 & 0.110\\
 \hline
\end{tabular}
\caption{List of the reference star id and the number of times that it is seen in total
and the values of its astrometric parameters (i.e. parallaxes and proper motions with their respective
errors) from Gaia DR2.}
\label{table:1}
\end{table*}

\begin{table*}
\centering
\begin{tabular}{| c |c c c c c c c c c c c c c c c c c c |} 
 \hline
 transit no. & 1 & 2 & 3 & 4 & 5 & 6 & 7 & 8 & 9 & 10 & 11 & 12 & 13 & 14 & 15 \\ 
 no. of ref stars & 13 &  8 & 20 & 18 & 31 & 25 & 27 & 31 & 23 & 28 & 22 & 22 & 28 & 21 & 28 \\
 \hline
\end{tabular}
\caption{Total number of reference stars seen on each transit.}
\label{table:2}
\end{table*}

\subsection{Effect due to the AL-AC motion of a star}\label{sect:alacrate}
We need to define a set of frames (our 'plates'), each one identified by a unique observing time. 
For convenience we can choose it to be the $t_{ot}$ of the target star - 
to which the measurements of all the stars observed some $\pm\Delta t$ from $t_{ot}$ are referred to.
For each $t_{ot}$ we will have a different local frame, with coordinate axes identified by the along 
and across scan direction at time $t_{ot}$, and we must model accordingly all measurements that we 
assign to that frame.
Then, we use the principles of differential astrometry to adjust all the frames to a common frame by 
means of translations, rotations, scale terms and possible distortion terms if necessary.

It is important that for each star observed at time $t_{os}$, such that
\begin{equation} 
|(t_{ot}-t_{os})| < \Delta t
\end{equation}
we refer its measured coordinates to the same local frame. To do so, we first compute the star's field
angles ($\eta,\zeta$) using the best current calibration parameters; 
then, we use the time derivatives of the field angles ($d\eta/dt,d\zeta/dt$) to take into account the 
variation of the direction of the ($\eta$,$\zeta$) axes that occurred in the elapsed time $\Delta t$.
To first order approximation, the local coordinates of a star observed in the frame defined by $t_{ot}$ are: 
\begin{align}
  \eta(t_{ot})  &= \eta(t_{os})+{\dot\eta}\times(t_{ot}-t_{os}) \nonumber \\
  \zeta(t_{ot}) &= \zeta(t_{os})+{\dot\zeta}\times(t_{ot}-t_{os})
\end{align}		         

For the field-angle rates we can use the formulae: 
\begin{align}
  \dot\eta &= -\omega_z + [\omega_x \cos\varphi + \omega_y \sin\varphi] \tan\zeta \nonumber \\
  \dot\zeta &= - \omega_x \sin\varphi + \omega_y \cos\varphi
\end{align}
where $\omega_x, \omega_y$ and $\omega_z$ are the components of the instantaneous inertial angular 
velocity of Gaia, and $\varphi = \eta \pm \gamma/2$, where the plus or minus sign is used for 
preceding or following FOV respectively and $\gamma$ being the basic angle between the
two fields of view of Gaia.

\subsection{Calibration effects}\label{sect:calibrations}

Amongst the possible distortion effects
we take into account the Large-scale AL and AC calibrations where the former 
could potentially vary over time scales of a day.
As we are considering observations over a couple of days, we 
assume these large scale calibrations to be constant to first approximation.
The AL large-scale calibration is modeled as a low-order polynomial in the across-scan pixel coordinate 
$\mu$ (that varies from 13.5 to 1979.5 across the CCD columns, \citealt{lindegren2012}) and can be written as:
\begin{equation}\label{eqn:calibe}
\eta_{fn}(\mu,t) = \eta^0_n + \sum_{r=0}^2 \Delta\eta_{rfn} L_r^*\left(\frac{\mu-13.5}{1966}\right)  
\end{equation}
where \textsl{f} is the FOV index, \textsl{n} is the CCD index and \textsl{r} is the degree 
of the shifted Legendre polynomial $L_r^*(\tilde{\mu})$ as a function of the normalized AC pixel coordinate 
($\tilde{\mu}$) and $\eta^0_n$ is the nominal calibration.
A similar equation holds for the AC large-scale calibrations which can be written as:
\begin{equation}\label{eqn:calibz}
\zeta_{fn}(\mu,t) = \zeta^0_{fn} + \sum_{r=0}^2 \Delta\zeta_{rfn} L_r^*\left(\frac{\mu-13.5}{1966}\right)  
\end{equation}

We will assume non-gated observations which is technically only valid for faint sources; brighter star
observations involve as many as a dozen gates that would need to be calibrated. 
Saturated images by stars brighter than about G = 12  is
mitigated through the use of TDI gates, activated to inhibit charge transfer
and hence to effectively reduce the integration time for bright sources \citep{brown2016}. The adopted TDI gate scheme impacts the bright-star errors and it can be seen that predicted end-of-mission parallax errors averaged over the sky 
vary in the range of 5-16 $\mu$as (see Fig. 10 in \citealt{deBruijne2012}).
However, the advantage of using sources brighter than G $\sim$ 13 mag is that they will always be 
observed as two-dimensional images, 
and therefore have high-precision measurements in both coordinates, though the precision of the AC coordinate is a factor of five worse than the AL one due to the rectangular shape of the CCD pixels (\citealt{debruijne2005}, also see Table 1 in \citealt{abbas2017}).

\section{Transformation and fitting procedure}\label{sect:transforms}
The field angles are transformed by first rotating them onto the (1,0,0) vector and then projecting 
them onto the tangent plane through a gnomonic transformation. The least squares fitting procedure 
then involves the taylor series expansion of the transformed field coordinates.

\subsection{A priori removal of parallaxes and proper motions}\label{sect:apriori}
With the aim of keeping the number of unknowns to be estimated to a minimum, we show the procedure 
of removing a priori the parallaxes and proper motions from Gaia DR2. 
As can be seen in Table\ref{table:1} the Gaia DR2 proper motions for this set of stars varies up to 116 mas/yr and with parallaxes up to 9 mas. In the simulation we account for these errors as gaussian distributions with standard deviations from Gaia DR2.
These quantities are given in equatorial coordinates that then need to be converted to the Satellite 
Reference System (SRS) giving the projected along and across scan values. 
This transformation is completely determined by the position angle of the scan given by:
\begin{align} \label{eqn:transform_scan}
    \mu_\eta &= \mu_\alpha \sin \theta + \mu_\delta \cos \theta
  \nonumber \\
    \mu_\zeta &= - \mu_\alpha \cos \theta + \mu_\delta \sin \theta
\end{align}
When treating the parallaxes, $\mu_\alpha$ and $\mu_\delta$ can be replaced with the parallax
factors $f_\alpha$ and $f_\delta$ respectively to obtain the transformed parallax factors along 
and across the scanning direction.
These transformed quantities are then subtracted from the $\eta$ and $\zeta$ positions of the stars
after multiplying the proper motions by the time passed since a chosen reference epoch and the
parallaxes by the parallax factors. \\
The 'corrected' field angles are then given as:
\begin{align} \label{eqn:apriori}
    \eta &= \eta_{calc} - \mu_\eta(t-T) - f_\eta \pi
  \nonumber \\ 
    \zeta &= \zeta_{calc} - \mu_\zeta(t-T) - f_\zeta \pi
\end{align}
where $\eta_{calc}$ and $\zeta_{calc}$ are calculated from the CCD and FOV information such as the
CCD row and column numbers and FOV index combined with the geometrical calibrations. The proper direction to 
the star at a given time is then obtained through the field angles by including the attitude at that instant. 

\subsection{Pre-rotation and Gnomonic transformation}\label{sect:gnomonic}
The fundamental ingredients to the model are the SRS field angles ($\eta$, $\zeta$) 
for these stars including their geometric calibrations (Sec.\ref{sect:calibrations}) and AL-AC motion (Sec.\ref{sect:alacrate}). Furthermore, the operations mentioned in Sec.\ref{sect:apriori} 
are performed that involve the estimation and a priori removal of 
the Gaia DR2 proper motions and parallaxes. We denote the target star as measured in the 
$i$th frame by $\eta^{'}_{i0}$ and $\zeta^{'}_{i0}$, whereas $\eta^{'}_{ij}$ and $\zeta^{'}_{ij}$ are the observed calibrated field angles of the $j$th star in the $i$th frame. 

Before trying a global adjustment of all the frames using the principles of tangent-plane astrometry, field angle measurements must be rectified via a gnomonic transformation. 
In order to minimize the differential effect of the so-called $\it tilt$ terms, which are second-order quantities arising from a misalignment of the nominal vs. true 
position of the telescope's optical axis, all the reference stars, including the target, are rotated in such a way that the position of the target star becomes aligned with the (1,0,0) vector defining the optical axis of Gaia.

Then these field angles are converted to coordinates in the tangent plane with tangent point (0,0)
using gnomonic transformations as follows:
\begin{equation}\label{eqn:gnomonic}
  X^{'}_{ij} = \tan(\eta^{'}_{ij}), \quad 
  Y^{'}_{ij} = \frac{\tan(\zeta^{'}_{ij})}
  {\cos(\eta^{'}_{ij})}
\end{equation}
for the $i$th frame and the $j$th star.

\subsection{Taylor series expansion}\label{sect:taylor}
We linearize the expressions from eq.\ref{eqn:gnomonic} and perform a Taylor series expansion of the reference star coordinates in the tangent plane 
from above around the nominal position of the star, i.e. the fiducial position free from any 
calibrations. This operation has the added advantage of reducing the number of 
calibration unknowns by a factor of three with respect to the astrometric model
used in \citealt{abbas2017} as will be shown below.

The nominal positions will be denoted by the unprimed quantities $\eta_{ij}$ and $\zeta_{ij}$. 
The expansion to first order can be written as:
\begin{align} \label{eqn:taylor}
  X^{'}_{ij} &= X_{ij}\Bigr|_{\substack{\eta_{ij} \zeta_{ij}}} 
  + (\eta^{'}_{ij} - \eta_{ij}) \frac{\partial X'_{ij}}{\partial \eta} + 
        (\zeta^{'}_{ij} - \zeta_{ij}) \frac{\partial X'_{ij}}{\partial \zeta}
  \nonumber \\  
  Y^{'}_{ij} &= Y_{ij}\Bigr|_{\substack{\eta_{ij} \zeta_{ij}}} + (\eta^{'}_{ij} - \eta_{ij}) \frac{\partial Y'_{ij}}{\partial \eta} + 
        (\zeta^{'}_{ij} - \zeta_{ij}) \frac{\partial Y'_{ij}}{\partial \zeta}
\end{align}

where the partial derivatives are evaluated at the nominal position and are treated as known quantities.
They are given as:
\begin{align} \label{eqn:partialX}
  \frac{\partial X'_{ij}}{\partial \eta}\Bigr|_{\substack{\eta_{ij} \zeta_{ij}}} &= \frac{\cos(\zeta_{ij})\cos(\eta_{ij})}{f(\eta_{ij}, \zeta_{ij})} - \frac{\cos(\zeta_{ij})\sin(\eta_{ij})}{f^2(\eta_{ij}, \zeta_{ij})} \nonumber \\
  &\times \left(-\cos(\zeta_{ij})\sin(\eta_{ij})\right)
  \nonumber \\  
  \frac{\partial X'_{ij}}{\partial \zeta}\Bigr|_{\substack{\eta_{ij} \zeta_{ij}}} &= -\frac{sin(\zeta_{ij})sin(\eta_{ij}}{f(\eta_{ij}, \zeta_{ij})} - \frac{cos(\zeta_{ij})sin(\eta_{ij})}{f^2(\eta_{ij}, \zeta_{ij})} \nonumber \\
  &\times \left(- sin(\zeta_{ij})cos(\eta_{ij})\right)
  \nonumber \\  
\end{align}
\begin{align} \label{eqn:partialY}
  \frac{\partial Y'_{ij}}{\partial \eta}\Bigr|_{\substack{\eta_{ij} \zeta_{ij}}} &= 
  - \frac{\sin(\zeta_{ij})}{f^2(\eta_{ij}, \zeta_{ij})} \times \left(-\cos(\zeta_{ij})\sin(\eta_{ij})\right)
  \nonumber \\  
  \frac{\partial Y'_{ij}}{\partial \zeta}\Bigr|_{\substack{\eta_{ij} \zeta_{ij}}} &= -\frac{\cos(\zeta_{ij})}{f(\eta_{ij}, \zeta_{ij})} 
  - \frac{\sin(\zeta_{ij})}{f^2(\eta_{ij}, \zeta_{ij})} \nonumber \\
  &\times \left(- \sin(\zeta_{ij})\cos(\eta_{ij})\right)
  \nonumber \\  
\end{align}
where $f(\eta_{ij}, \zeta_{ij}) = 
\cos(\zeta^{'}_{ij})\cos(\eta^{'}_{ij})$.

The differences multiplying the partial derivatives are the calibration terms given in eqns.\ref{eqn:calibe} $\&$
\ref{eqn:calibz} that are to be estimated through the least-squares technique mentioned later on. 

\subsection{Astrometric model}\label{model}
The positions of the stars in each frame are adjusted to its reference frame value through a simple 
linear plate transformation given as:
\begin{align} \label{eqn:full_linPM}
  X^{'}_{0j}(t_{0t}) &= a_i X^{'}_{ij}(t_{it}) + b_i Y^{'}_{ij}(t_{it}) + c_i 
  \nonumber \\
 Y^{'}_{0j}(t_{0t}) &= d_i X^{'}_{ij}(t_{it}) + e_i Y^{'}_{ij}(t_{it}) + f_i 
\end{align}
where $X^{'}$ and $Y^{'}$ are the star
coordinates in the tangent plane, with $X^{'}_{0j}$ and $Y^{'}_{0j}$ being the coordinates of the $j$th star on the
reference frame.
The plate constants are given by $a_i$, $b_i$, $c_i$, $d_i$, $e_i$ and $f_i$.  

The software package GAUSSFit \citep{jefferys1988} is used to solve this set of equations through a least squares procedure that minimizes the 
sum of squares of the residuals weighted according to the input errors. 
The estimated plate/frame parameters ($a_i$ through $f_i$) allow to transport the calibrated observations ($X^{'}_{ij}$, $Y^{'}_{ij}$) to a common reference frame. The distribution of residuals then informs us as to how well the model
accounts for various effects.
As can be expected, it is found that $a_i$ and $e_i$ are almost unity,
whereas $b_i$ = $-d_i$  and together they give the rotation and orientation. 
The offsets $c_i$ and $f_i$ give the zero point of the common system (see \citealt{abbas2017}, for more details).

\section{Results}\label{results}
We summarize here the most relevant findings of our numerical experiments aimed at determining the ultimate  systematic floor of our methodology for differential astrometry. We study in particular its behaviour a) as a function of unmodelled parallaxes and proper motions that are not removed a priori, b) when taking into account catalog errors in the astrometric parameters, and, c) in the case of lower-precision knowledge of the satellite 
velocity.

\subsection{Systematic floor of the methodology}\label{sec:sysfloor}
\begin{figure*}
    \centering
    \includegraphics[width=0.98\textwidth]{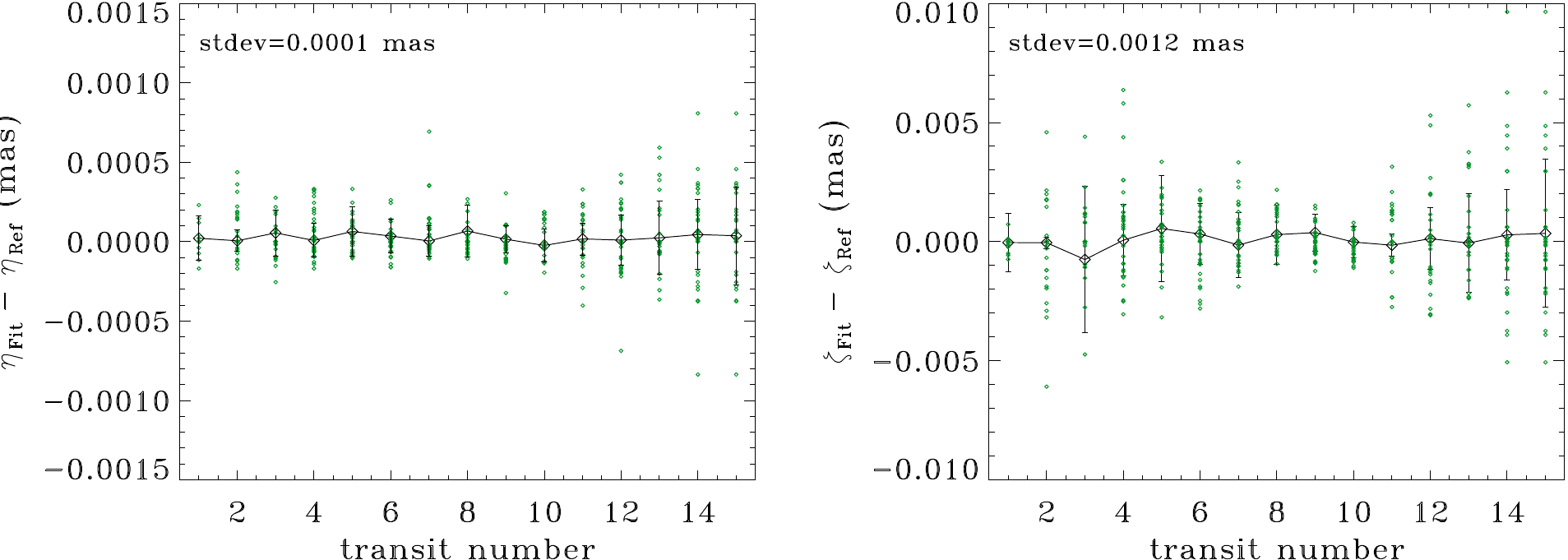}
    \caption{The residuals after removing a priori the proper motions and parallaxes of the 
    stars as a function of the transit number with no added physical effects 
    demonstrating the accuracies at the sub$\mu$as-level in AL (left panel) and 
    $\mu$as-level in AC (right panel). 
    The points show the average differences of the best-fit positions for a 
    reference star and its position on the reference frame on any given transit, where the 
    green points are the mean differences for the star on that transit and the black points 
    and error bars depict the average over all the stars for that transit and their standard
    deviation respectively. 
    }
    \label{fig:remPMsPar}
\end{figure*}

\begin{figure*}
    \centering
    \includegraphics[width=0.98\textwidth, trim=5mm 20mm 20mm 90mm, clip]{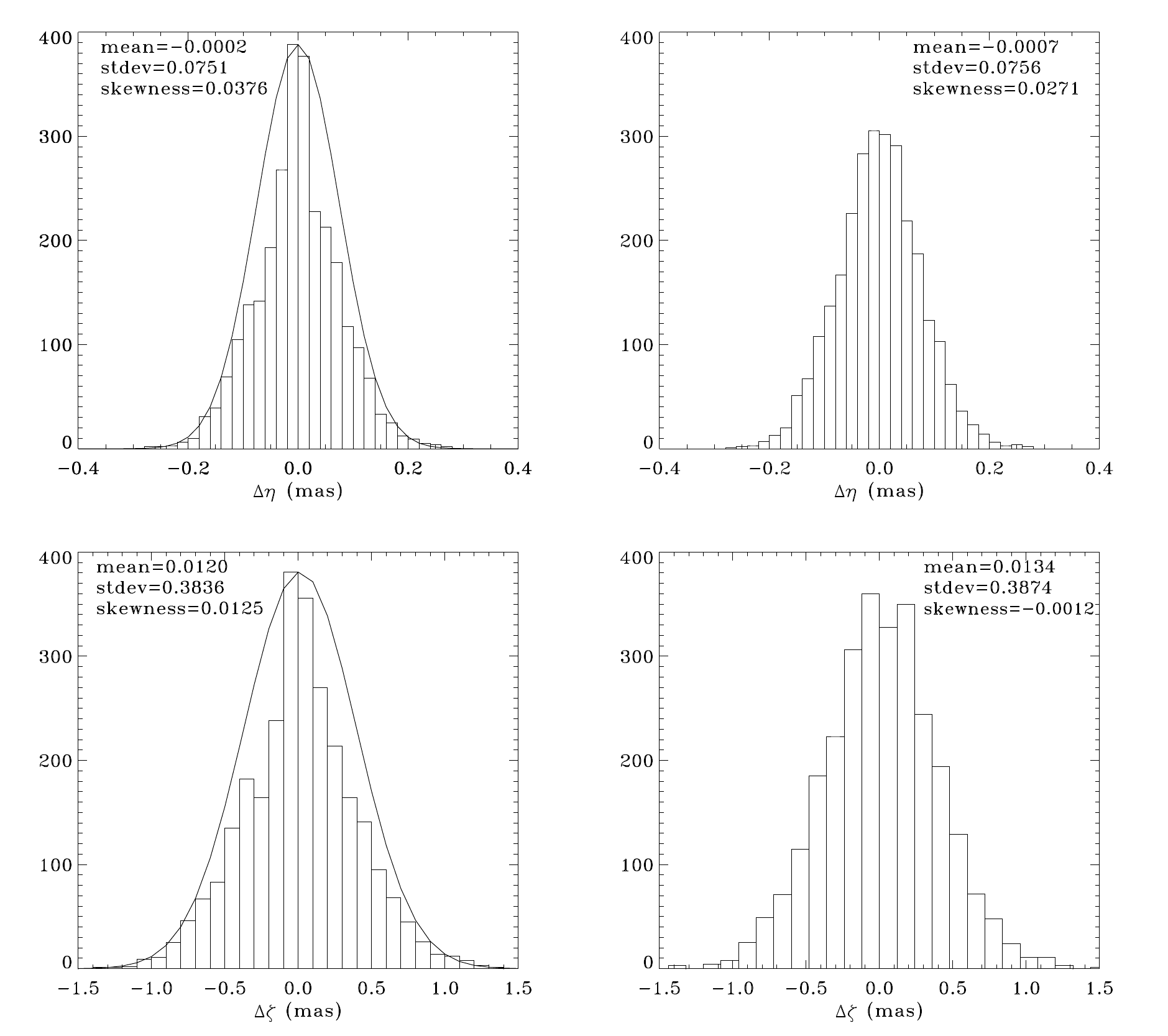}
    \caption{The residuals after removing a priori the proper motions and parallaxes of the stars  (left panels) with input gaussian errors (right panels) that follow the standard magnitude-dependent errors \citep{abbas2017}.}
    \label{fig:wGaussErrs}
\end{figure*}

Fig.\ref{fig:remPMsPar} shows the sub$\mu$as-level stability 
achievable in the AL direction (1$\mu$as in AC)
when no physical effects are included in the simulation (i.e. aberration and gravitational 
light deflection are turned off) and after an a priori removal of 
the star's parallax and proper motion (following the procedure 
mentioned in Sec.\ref{sect:apriori}).
The figure shows the differences between the fitted $\eta$ and $\zeta$ per star (Eq.\ref{eqn:full_linPM}) using the best-fit plate and calibrations parameters 
obtained through the least-squares procedure of GAUSSFit and the corresponding values on the reference frame as a function of the transit number.
Each transit is the average over eight frames from the first eight columns in the astrometric field (reason for this is described at the end of Sec.\ref{sect:refstars}),
and the green points depict the average differences with 
respect to the reference frame values for each star seen on that transit. 
The black points are the averages over all the reference stars for that transit and the error bars show the standard deviation of the green points.
The AL/AC input error and output residual distributions are shown in Fig.\ref{fig:wGaussErrs} demonstrating the full recovery of the input standard errors in the methodology.

\subsection{Residuals due to unmodeled effects}
\begin{figure*}
    \centering
    \includegraphics[width=0.98\textwidth]{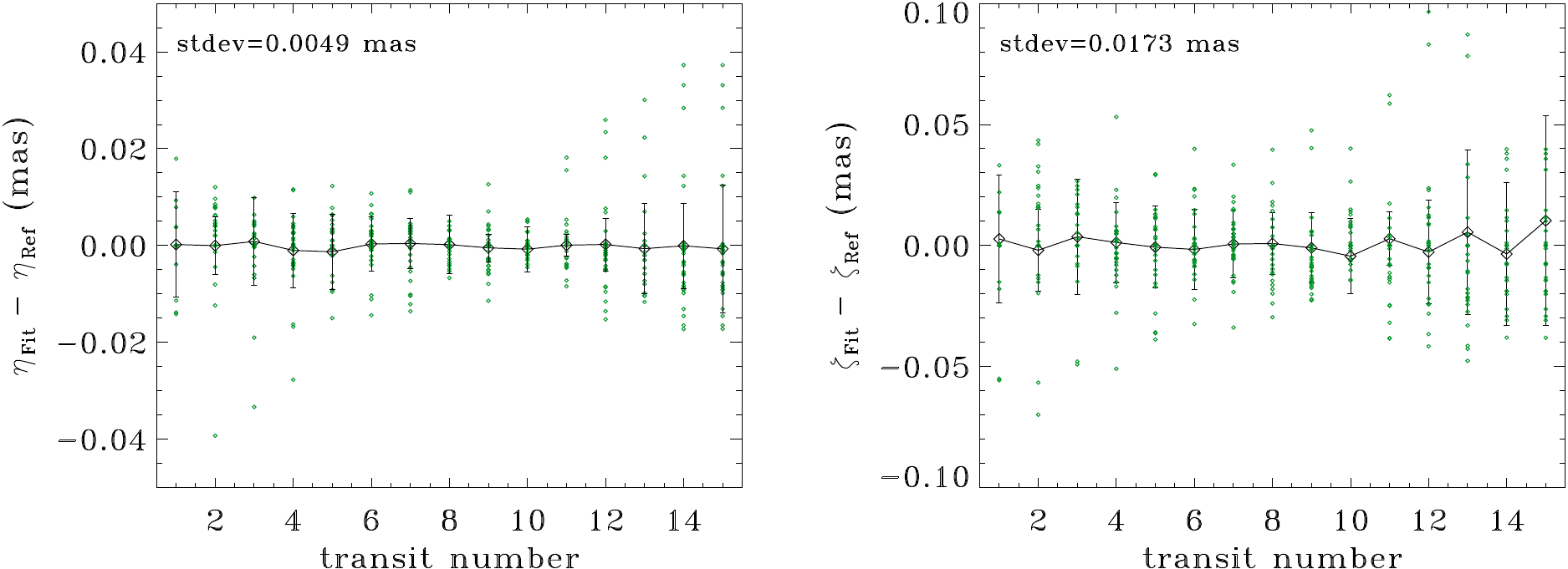}
    \caption{The stability of reference stars with unmodeled Parallaxes 
    that do not have their parallax estimates removed 
    a priori (only proper motion estimates are removed). The left panel shows the AL differences in the 
    standard deviations and the right panel shows the AC differences.}
    \label{fig:unmodPar}
\end{figure*}

\begin{figure*}
    \centering
    \includegraphics[width=0.98\textwidth]{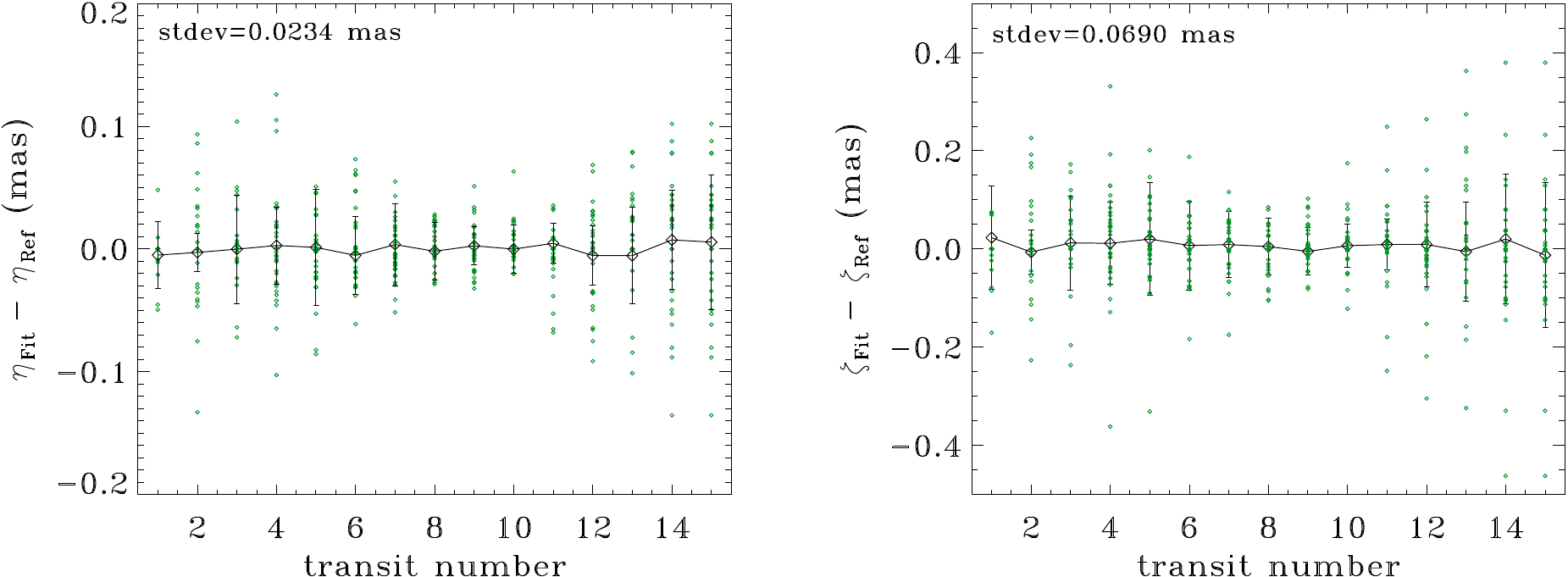}
    \caption{Same as Fig.\ref{fig:unmodPar} but now for unmodeled proper motions
    and parallaxes estimates removed a priori.
   }
    \label{fig:unmodPMs}
\end{figure*}

Here we will look at the effect of not accounting for the stars' parallaxes nor their 
proper motions by neither removing their estimated values a priori nor attempting 
to model their contribution with appropriate unknown parameters that are subsequently
estimated through a least-squares fitting procedure. 
These astrometric parameters would then augment the systematic floor and 
Figs.\ref{fig:unmodPar} and \ref{fig:unmodPMs} show the contributions from unmodelled 
parallaxes and proper motions respectively to the differences between the best-fitted 
star parameters for that transit and its reference frame value (see the previous Sec.\ref{sec:sysfloor} for a more detailed description of the plot and points).

We can see that unmodeled proper motions have a 4-5 times larger effect on the stability than unmodeled parallaxes, where unmodeled Gaia DR2 proper motions introduce extra
residuals of $\sim$23$\mu$as (AL) and 69$\mu$as (AC) versus the $\sim$5$\mu$as (AL) and 17$\mu$as (AC) due to unmodeled parallaxes.
This result stems from the contribution of the average proper motions of the field (observed over the short time scale of the GAREQ experiment) and from that due to the effective parallax offset of the full field. 
It can also be argued that the smaller magnitudes of the parallaxes versus the much larger proper motions in combination with the time of year (contributing to the size of the parallax factor) and time span of observations are critical factors as well.


\subsection{Effect of parameter errors}

\begin{figure*}
    \centering
    \includegraphics[width=0.98\textwidth]{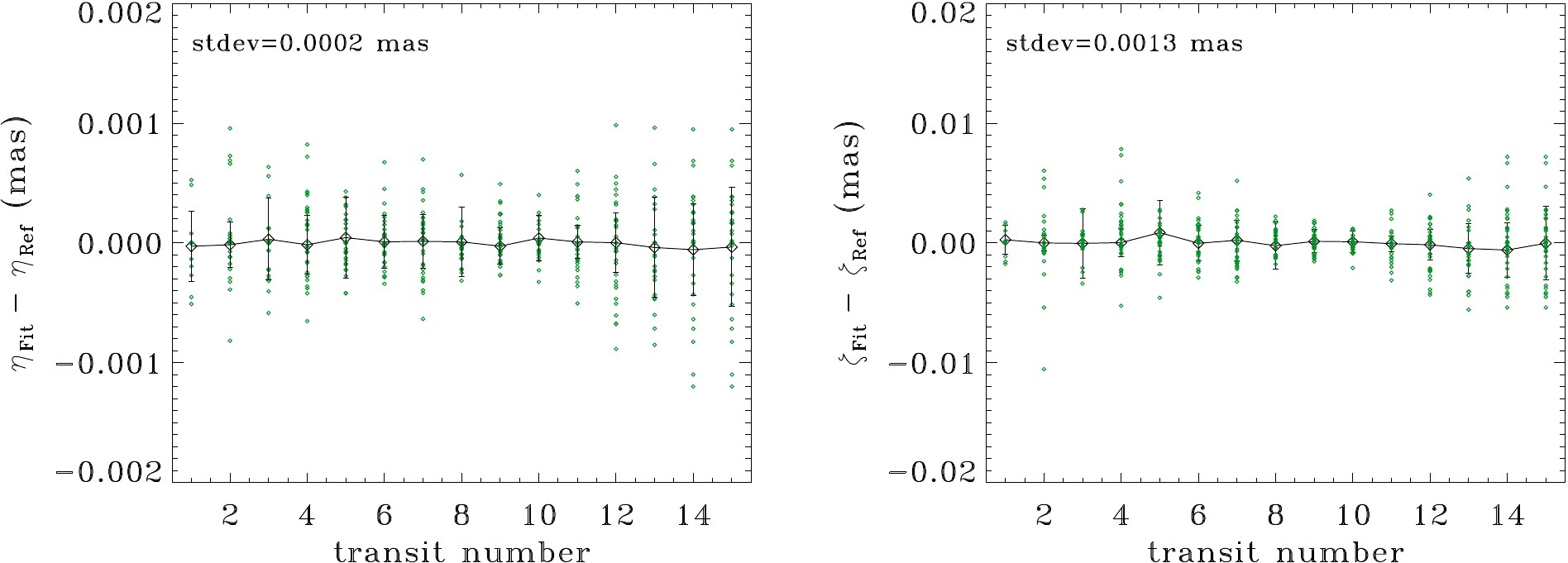}
    \caption{The stability of reference stars that have Gaia DR2 errors in their parallaxes.
    The left panel shows the AL differences in the 
    standard deviations and the right panel shows the AC differences.}
    \label{fig:Parwerrs}
\end{figure*}

\begin{figure*}
    \centering
    \includegraphics[width=0.98\textwidth]{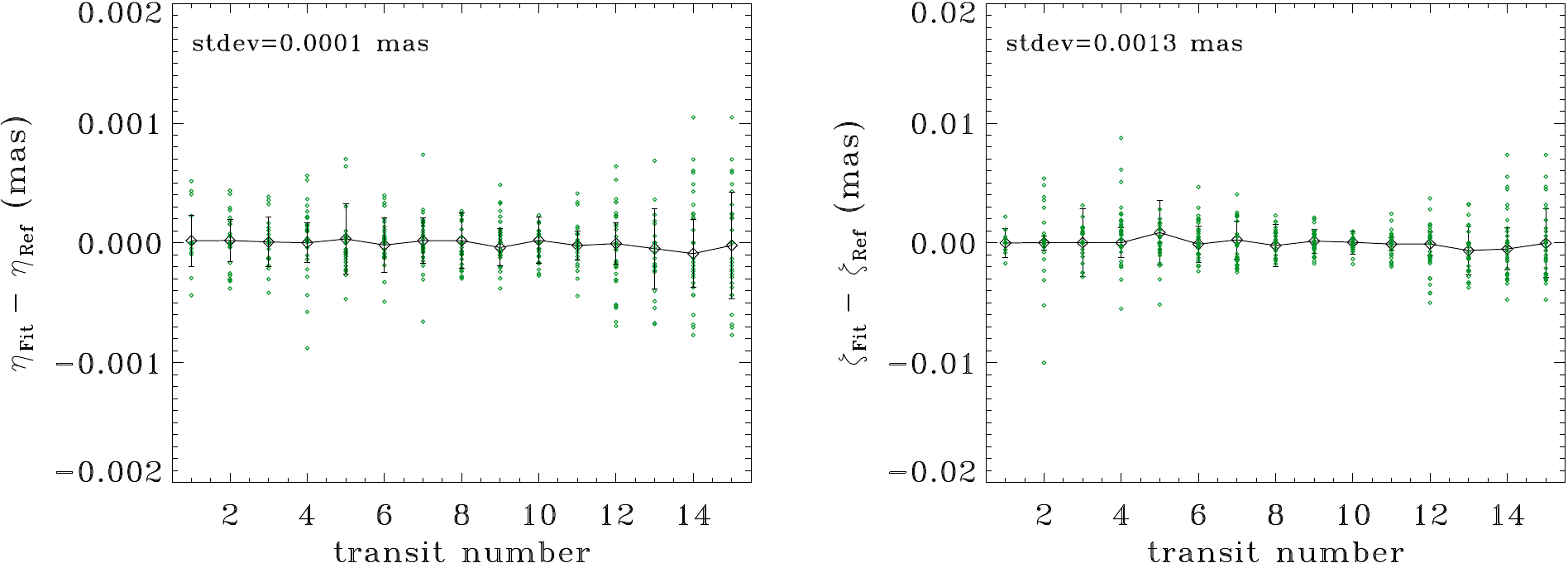}
    \caption{The stability of reference stars with errors in their proper motions 
    as given by the Gaia DR2 catalog.}
    \label{fig:PMswerrs}
\end{figure*}

In order to see the effect that catalog errors in the astrometric parameters have on the 
stability, we add them as gaussian deviates to the parameter values and remove projected  
estimates of the original error-free astrometric parameters a priori.
Figs.\ref{fig:Parwerrs} and \ref{fig:PMswerrs} show the effects of the catalog errors in the parallaxes and proper motions. We can see that the 
stability remains at the sub $\mu$as-level in AL and $\mu$as level in AC for 
both the parallaxes and proper motions demonstrating the minimal effect of such 
errors.

\begin{figure*}
    \centering
    \includegraphics[width=0.98\textwidth]{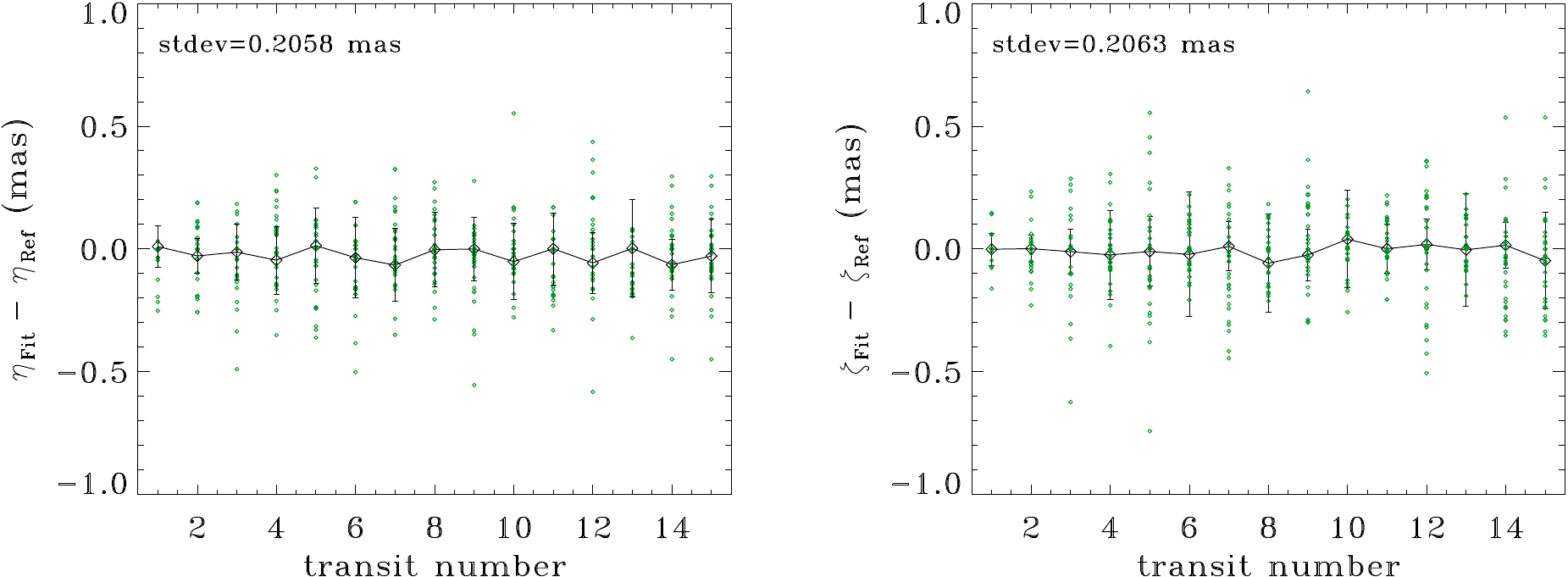}
    \caption{The stability of reference stars with 10$\mu$as/s errors in each of the
    components of the satellite velocity.}
    \label{fig:AlAcrate}
\end{figure*}

To see the effect an imprecise satellite velocity has on the residuals, we artificially perturb its value by adding a time-dependent error of 10$\mu$as s$^{-1}$ in each of the components of the satellite velocity. The effect is shown in Fig.\ref{fig:AlAcrate} giving AL and AC scan residuals of roughly 0.2mas each.
We noticed that by introducing even a small time dependence on the uncertainty quickly 
translates into a much larger error on the stability of the reference system over the time
span of the observations. This is a highly conservative estimate in light of a variety
of attitude errors such as microclanks and micrometeoroid hits \citep{lindegren2016}.
Interestingly enough, this type of error contributes large residuals eventually giving 
AL and AC stability values that are almost equal.


\section{Summary and discussion}\label{discussion}

In this paper we have further developed a differential astrometric framework for the analysis of Gaia observations, initially introduced in Abbas et al. (2017). We have specifically targeted high-cadence, short timescale observations in one of the fields of the GAREQ experiment, which is set to provide an unprecedented measurement of the relativistic deflection of light due to Jupiter's quadrupole and a first-time measurement in the optical of the light deflection due to Jupiter's monopole (expected to be 
$\sim$16mas for a grazing light, \citealt{crosta2006}) . 

We have demonstrated the sub$\mu$-as level stability of a local reference frame composed of a few tens of comparison stars surrounding the bright target star that is expected to show a large value for the relativistic light deflection due to its proximity to Jupiter's limb.
Our differential astrometric methodology relies upon pre-rotations that minimize second-order differential effects by transporting the various frames containing the field angles at different times to be overlapping around the target star, aligned with the unit vector defining the optical axis of Gaia. Subsequent gnomonic transformations convert the field angles into tangent plane coordinates centered on the target star. Furthermore, a Taylor series expansion of each reference star's position around its fiducial point allows us to keep the number of calibration parameters to be estimated at a minimum, i.e. only two per CCD for each direction.

We show that a procedure for removing a priori error-free parallaxes and proper motions of reference
stars brings the advantage of reducing the number of unknown parameters to be estimated to only the calibration and plate parameters. Indeed if such a procedure is not adopted and the parallaxes and proper motions are left unmodeled, the residuals are enlarged with the most dominant effect (by a factor of 4-5) being that due to the proper motion.
The inclusion of catalog errors in the parallaxes and proper motions 
has a minimal impact, i.e. at the sub-$\mu$as level in AL, mainly due to the smaller
magnitudes in the errors.
We also show that a lower precision knowledge of the satellite velocity with 
10$\mu$as s$^{-1}$ errors in its velocity components has a 
relatively important effect on the residuals giving conservative estimates
of $\sim$ 0.2mas in both AL and AC.

We plan next to apply the methodology presented here to the analysis of the actual observations of events in the  GAREQ experiment, and carry out detailed modeling of the various effects discussed in this paper. Additional, future applications of the differential astrometric approach to the analysis of Gaia observations naturally include configurations in which Gaia data are gathered on significantly longer timescales than those (around 24 hr) encompassed by the GAREQ experiment. Possible science cases include studies of future microlensing events (e.g., \citealt{mustill2018, bramich2018}), the astrometric detection of orbital motion induced by planetary-mass companions (e.g., \citealt{sozzetti2014, sozzetti2016, perryman2014}), and investigations of the brown dwarf binary fraction (e.g., \citealt{marocco2015}, and references therein). All the above analyses will require further improvements in the modeling of calibration and instrument attitude effects, that become important for Gaia astrometric time series with years-long baselines. A robust modeling framework for differential astrometry is also mandatory in view of future developments in the field aimed at pushing the ultimate precision in position measurements in the $\mu$as regime (e.g., \citealt{malbet2016}) significantly beyond that achievable by Gaia.


\section*{Acknowledgements}
We thank Alessandro Sozzetti for a careful reading of the paper and many
useful comments.  We also thank the anonymous referee for their highly constructive
report that has helped to improve the paper.
This work was supported by the Italian Space Agency through Gaia
mission contracts: the Italian participation to DPAC, ASI 2014-
025-R.1.2015 in collaboration with the Italian National
Institute of Astrophysics.




\bibliographystyle{mnras}
\bibliography{gareq} 








\bsp	
\label{lastpage}
\end{document}